\documentclass{article}

\usepackage[preprint, nonatbib]{neurips_2024}
\usepackage{array}

\usepackage[utf8]{inputenc} 
\usepackage[T1]{fontenc}    
\usepackage{hyperref}       
\usepackage{url}            
\usepackage{booktabs}       
\usepackage{amsfonts}       
\usepackage{nicefrac}       
\usepackage{microtype}      
\usepackage{xcolor}         
\usepackage{amsmath}

\usepackage{graphicx} 
\usepackage{wrapfig} 
\usepackage{booktabs} 
\usepackage{subcaption}
\usepackage{siunitx}
\usepackage{enumitem}  

\setlist[enumerate, 1]{left=5pt}  

\usepackage{todonotes} 

\newif\ifshowcomments

\definecolor{softred}{RGB}{255,99,71}
\definecolor{softblue}{RGB}{100,149,237}
\definecolor{softgreen}{RGB}{60,179,113}

\newcommand{\hf}{\mathfrak{h}}  

\ifshowcomments
  \newcommand{\lars}[1]{\textbf{\textcolor{softblue}{[Lars: ]}} \textcolor{softblue}{#1}}
    \newcommand{\jules}[1]{\textbf{\textcolor{sofgreen}{[Jule: ]}} \textcolor{softgreen}{#1}}
\else
  \newcommand{\lars}[1]{}
  \newcommand{\jules}[1]{}
\fi

\usepackage[backend=biber,style=nature,sorting=none]{biblatex}
\title{BoostMD: Accelerating molecular Sampling using ML Force Field Feature}
\title{BoostMD: Regressing changes in Force Field features for Accelerating Molecular Sampling}
\title{BoostMD: Accelerating Molecular Sampling \\ by leveraging ML force field features computed at previous time-steps}
\title{BoostMD: Accelerating molecular sampling \\ by leveraging ML force field features from previous time-steps}

\author{%
  Lars L. Schaaf \\
  University of Cambridge \\
  InstaDeep \\ 
  \texttt{lls34@cam.ac.uk}
  \And
  Ilyes Batatia \\
  University of Cambridge\\
  \texttt{ib467@cam.ac.uk}\\
  \And
    Christoph Brunken \\
  InstaDeep \\
  \texttt{c.brunken@instadeep.com} \\
  \And
  Thomas D. Barrett \\
  InstaDeep \\
  \texttt{t.barrett@instadeep.com} \\
  \And 
  Jules Tilly\\
  InstaDeep\\
  \texttt{j.tilly@instadeep.com}\\
}

\begin{document}

\maketitle

 \vspace{-.4cm}
\begin{abstract}

    Simulating atomic-scale processes, such as protein dynamics and catalytic reactions, is crucial for advancements in biology, chemistry, and materials science. Machine learning force fields (MLFFs) have emerged as powerful tools that achieve near quantum mechanical accuracy, with promising generalization capabilities. However, their practical use is often limited by long inference times compared to classical force fields, especially when running extensive molecular dynamics (MD) simulations required for many biological applications. In this study, we introduce BoostMD, a surrogate model architecture designed to accelerate MD simulations. BoostMD leverages node features computed at previous time steps to predict energies and forces based on positional changes. This approach reduces the complexity of the learning task, allowing BoostMD to be both smaller and significantly faster than conventional MLFFs. During simulations, the computationally intensive reference MLFF is evaluated only every $N$ steps, while the lightweight BoostMD model handles the intermediate steps at a fraction of the computational cost. Our experiments demonstrate that BoostMD achieves an eight-fold speedup compared to the reference model and generalizes to unseen dipeptides. Furthermore, we find that BoostMD accurately samples the ground-truth Boltzmann distribution when running molecular dynamics. By combining efficient feature reuse with a streamlined architecture, BoostMD offers a robust solution for conducting large-scale, long-timescale molecular simulations, making high-accuracy ML-driven modeling more accessible and practical.

\end{abstract}

 \vspace{-.2cm}
\section{Introduction}
 \vspace{-.2cm}

Accurate modeling of atomic-scale processes, such as protein folding, catalysis, and carbon capture, is a long-standing challenge in computational biology, chemistry, and materials science~\cite{kovacs_mace-off23_2023, behler_generalized_2007, jumper2021highly, chen2020computational, }. 
Recent advances in geometric deep learning have led to the development of machine learning force fields (MLFFs), which can predict atomic forces and energies based on 3D atomic configurations at near quantum mechanical accuracy~\cite{behler_generalized_2007, bartok_gaussian_2010, schutt_schnet_2017,shapeev_moment_2016,bochkarev_efficient_2022,nigam_recursive_2020,batzner_e3-equivariant_2022, batatia_mace_2023}. MLFFs, also known as machine-learning interatomic potentials, promise cost-effective atom-scale insight into biomolecular structure, chemical reaction mechanisms, and materials properties, allowing for \textit{in-silico} screening and design of new drugs, catalysts and carbon capture devices~\cite{hannagan2021first, schaaf2023accurate, rhodes202417o, zhou2023device}. These models, trained on high-quality quantum mechanical data, typically scale linearly with system size, and are more than 3 orders of magnitude faster for even small systems, while maintaining predictive accuracy~\cite{batatia_mace_2023, batzner_e3-equivariant_2022}. 
The emergence of generalizable MLFF foundation models capable of modeling a wide range of molecules and materials represents a significant breakthrough~\cite{kovacs_mace-off23_2023, batatia_foundation_2024,}. Their transferable nature allows for direct application without a need to generate large, expensive datasets or train from scratch. Furthermore, although trained on only small molecules, they are sufficiently transferable for stable and accurate simulations of large peptides~\cite{kovacs_mace-off23_2023}.

However, the main limitation of modern MLFFs is not their accuracy, but their inference time. Sampling algorithms such as Markov Chain Monte Carlo or Molecular Dynamics (MD) simulations need small time steps on the order of femtoseconds (\SI{e-15}{\second}), to ensure reasonable acceptance rates and numerical stability, respectively. Sampling equilibrium configurations often requires hundreds of millions of such steps to capture relevant conformational changes over longer timescales. With inference speeds on the order of $10^6$ evaluations per day~\cite{kovacs_mace-off23_2023},\textit{} MLFFs remain computationally prohibitive for many large-scale biological and material science applications.

This work introduces BoostMD, an approach for accelerating molecular dynamics (MD) with machine learning force fields. For most equivariant architectures, the majority of computational expense in MLFFs arises from generating expressive, atom-centered features of the 3D environment. These features are used to predict energy and forces using relatively simple and computationally inexpensive read-out functions. At each step of a molecular dynamics run these features are computed from scratch for the new atomic environment. Our new architecture, BoostMD, leverages node features computed at previous time steps to predict energies and forces for subsequent configurations, based on change in position.  Using the information of the reference atom features, it is possible to train a much smaller and faster model to predict energy changes with minimal loss of accuracy. At inference the full reference MLFF model is evaluated only at every $N$ steps, while the faster BoostMD model operates in between. By leveraging previously computed features, a shallow and fast architecture remains accurate during these interim steps, significantly reducing the overall computational cost while maintaining accuracy. 

Our contributions are as follows:

\begin{itemize}
\setlength\itemsep{0em}  
  \item We introduce, BoostMD, \textbf{a new approach for accelerating} ML force fields, by using previously computed node features to predict the energy and forces on atoms of subsequent simulation steps.
  \item We describe the \textbf{design space} of possible BoostMD models, for both training and sampling. We empirically test and contrast various fundamental architectural choices. 
  \item We present the first BoostMD model, which is more than eight times faster than the reference model. We show BoostMD models are \textbf{transferable} to unseen molecules and able to accurately sample equilibrium structures with the right probability distribution. 
\end{itemize}

\section{Motivation}

The hidden features of large atomistic foundation models are highly expressive low-dimensional representations of 3D atomic environments. Their expressivity has been leveraged for zero-shot molecular generation~\cite{elijosius_zero_2024}, performing on par with purpose-trained diffusion models. Furthermore, the hidden features from models trained only on forces and energies have been used to directly predict other properties, such as the behavior of nuclear spins under external fields, with shallow multi-layer-perceptron readouts~\cite{ivkovic2024transfer}. 

\begin{figure}[tb]
    \centering
    \includegraphics[width=1\textwidth]{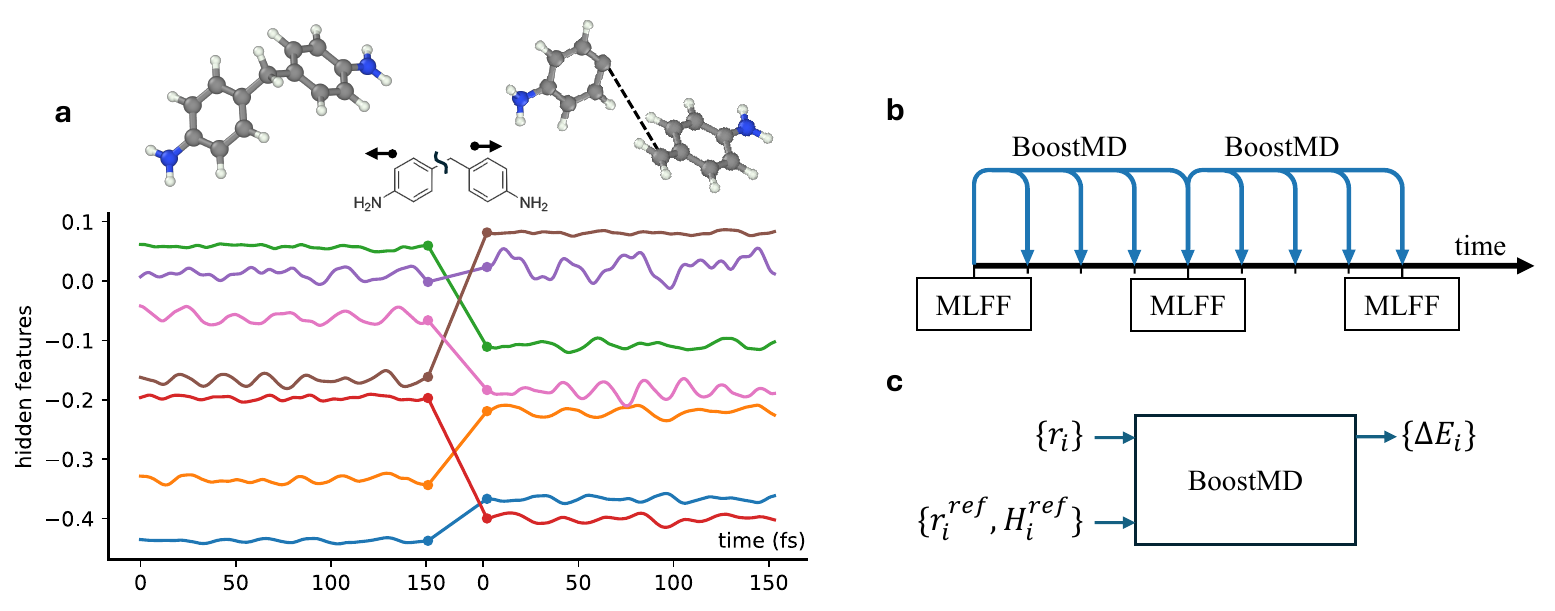}
    \caption{\textbf{Properties of hidden features and the BostMD architecture.} (\textbf{a}) Showing the node features of a foundation model for organic molecules (MACE-OFF23~\cite{kovacs_mace-off23_2023}) as a function of MD steps. After 150~fs the molecule is artificially split in two. The hidden features oscillate during MD and only significantly change under this sever configurational change. (\textbf{b}) Proposed method of using node features from previous time steps to predict energy differences using BoostMD models (\textbf{c}). }
    \label{fig:motivation-boost}
\end{figure}

During molecular dynamics, expressive node features are computed at each step but change minimally over time. Figure ~\ref{fig:motivation-boost}a shows how the node features oscillate during MD steps, with minimal drift or change. Only by artificially pulling the molecule apart do the node features qualitatively change. Continuing to run molecular dynamics on the two fragments, the node features of the separated molecule change minimally. 

Boost molecular dynamics (BoostMD) leverages the fact that node features at previous time steps are highly relevant for subsequent dynamics. 
Rather than discarding expressive features at every MD step, previously computed features can be used to help predict the energy of later steps, as visualized in Figure\ref{fig:motivation-boost}. This simplified task can be performed with smaller and faster architectures. 
Furthermore, the boost model can be restricted to be more local and depend only on the closest atoms. For large simulations with many atoms, this would allow BoostMD to be parallelized over multiple GPUs with less overhead cost, allowing for the scaling of atomistic simulations. 

\section{Related Work}

    \paragraph{Enhanced sampling techniques} Rare events, such as reactions, are challenging to observe with direct molecular dynamics, even when using very fast traditional force fields. Enhanced sampling methods introduce controlled biases to overcome energy barriers and access rare events. Samples can be reweighted to recover unbiased equilibrium observables. When combined with BoostMD, enhanced sampling methods amplify efficiency gains, as demonstrated in our experimental results, significantly increasing their practical applicability. Prominent enhanced sampling approaches include metadynamics~\cite{bussi_using_2020} and umbrella sampling~\cite{kastner_umbrella_2011}.
    
    \paragraph{Reference system propagator algorithms} In traditional non-ML force fields it is possible to evaluate a part of the force field every time step, while other, typically long-range terms of the force field are only evaluated every number of time steps~\cite{berne_molecular_1999,tuckerman_reversible_1992}. The core idea is that some terms (eg. short-range bonded interactions) change more rapidly than others (long-range electrostatics). Recently, a similar approach has been applied to ML force fields~\cite{fu_learning_2023}. This approach is different from BoostMD, as it does not bias the potential on the node features of previous time steps. BoostMD's speedup is independent of the interaction range and can be combined with such range-separated efforts. 
    \paragraph{Generative modeling} Generative models, including diffusion models and normalizing flows~\cite{tabak_family_2013}, offer an alternative to sequential approaches by directly generating samples from a target distribution~\cite{noe_boltzmann_2019, klein_transferable_2024, midgley_flow_2023,midgley_se3_2024, coretti_boltzmann_2024, rotskoff_sampling_2024, elijosius_zero_2024}. Although promising developments have been made in recent years, these models often lack transferability, require extensive data generation for each specific problem, or have only demonstrated effectiveness on small systems with lower dimensionality~\cite{noe_boltzmann_2019, klein_transferable_2024}. Consequently, they have yet to surpass the performance of MLFFs combined with traditional sampling methods. Unlike BoostMD, it is not trivial to combine these approaches with enhanced sampling techniques~\cite{klein_timewarp_2023}. 



\section{Background and Notation}

 \vspace{-.2cm}

\paragraph{Equivariant message-passing neural networks} Recent progress in equivariant message-passing neural networks~\cite{batzner_e3-equivariant_2022,batatia_mace_2023}, have allowed for unprecedentedly accurate MLFFs, also referred to as machine-learned interatomic potentials. In these models, a graph is constructed from a point cloud of atoms by connecting atoms within a cutoff distance. For equivariant models, the messages passed between nodes (atoms) encode the atom's geometric environment and transform with rotations in a controlled way. After a set number of message-passing steps, the node features pass through a readout that predicts the atom-centered energy (a scalar quantity). 
The majority of the computational cost originates from creating many-body, descriptive node features. The final readouts are small MLPs or linear layers~\cite{batatia_mace_2023, behler_generalized_2007, batzner_e3-equivariant_2022}. 
\vspace{-.2cm}

\paragraph{Irreducible representations and tensor products}
\label{sec:background-notation}
When learning equivariant functions from atomic positions, it is helpful to use the basis for irreducible representations of the $\text{SO}(3)$ group. The spherical harmonics $Y_m^l$, where $l$ is the degree and $m$ is the order, form a basis for all square-integrable functions on the unit sphere. The set of these functions forms a vector space on which group elements of SO(3), corresponding to rotations, can be represented. The effect of rotation on a learned function expanded as a linear combination of spherical harmonics is trivially given by the Wigner-D matrices. Node features expressed in the $Y_m^l$-basis can have scalar features ($l = 0$) that are invariant under rotation or higher-order features ($l > 0$), such as vectors, that transform equivariantly with rotations. 

In non-geometric machine learning, we can use tensor products of two features $(A\otimes B)_{ijkl} =A_{ij} B_{jk}$ to generate higher-dimensional tensors. For E(3) equivariant features, we need to map the output back to the basis for irreducible representation (irreps), allowing us to separate invariant from equivariant functions. By doing so, the coefficients for the weights are restricted as to maintain the underlying equivariance. Spherical tensors, like the spherical harmonics, are indexed by degree $l$ and order $m$. Assuming two spherical tensors ${A}_{l_1}^{m_1}$ and $ {B}_{l_2}^{m_2}$ a learnable fully connected tensor product is defined as
\begin{equation}
    D^{m_3}_{l_3} = (\boldsymbol{A} \underset{}{\overset{\boldsymbol{\alpha}}{\otimes}} \boldsymbol{B})^{m_3}_{l_3} :=  \sum_{l_1 m_1, l_2 m_2} C_{l_1 m_1, l_2 m_2}^{l_3 m_3} F_{l_1 l_2 l_3}\left(\boldsymbol{\alpha}\right) {A}_{l_1}^{m_1}  {B}_{l_2}^{m_2} ,
    \end{equation}
    where $F_{l_1 l_2 l_3}$ is a multi-layer perceptron (MLP) with scalar inputs $\boldsymbol{\alpha}$, $C_{l_1, m_1; l_2, m_2}^{l_3 m_3}$ are the Clebsch–Gordan coefficients and the resulting tensor $D^{m_3}_{l_3}$ is expanded in the basis of the irreducible representation of SO(3).
This is, for example, how the position is combined with the sending node features in equivariant MPNNs. For more details please see the \textit{e3nn} paper~\cite{geiger_e3nn_2022} or classic literature~\cite{bourbaki_lie_1994}. 

 \vspace{-.1cm}

\section{BoostMD}

 \vspace{-.1cm}

  \subsection{Architecture}
  \label{sec:architecture}
  The BoostMD model predicts the energy difference between the current configuration and a reference configuration with the help of the reference configuration's node features. In practice, the reference configurations are taken from previous molecular dynamics steps. Take an atom $i$, which was previously at the position $\mathbf{r}_i^{ref}$ with corresponding reference node features $\mathbf{H}^{ref}_i$ and local atom energy $E_i^{ref}$. The boost model predicts the change in energy $\Delta E_i$, based on the atoms within a cut-off distance $j \in \mathcal{N}(i)$, including their current $\{\mathbf{r}_j \}$ and previous positions $\{\mathbf{r}^{ref}_j \}$ and corresponding previous node features $\{\mathbf{H}^{ref}_j \}$. 

\begin{figure}[tb]
    \centering
    \begin{subfigure}[b]{0.95\textwidth}
        \centering
        \includegraphics[width=\textwidth]{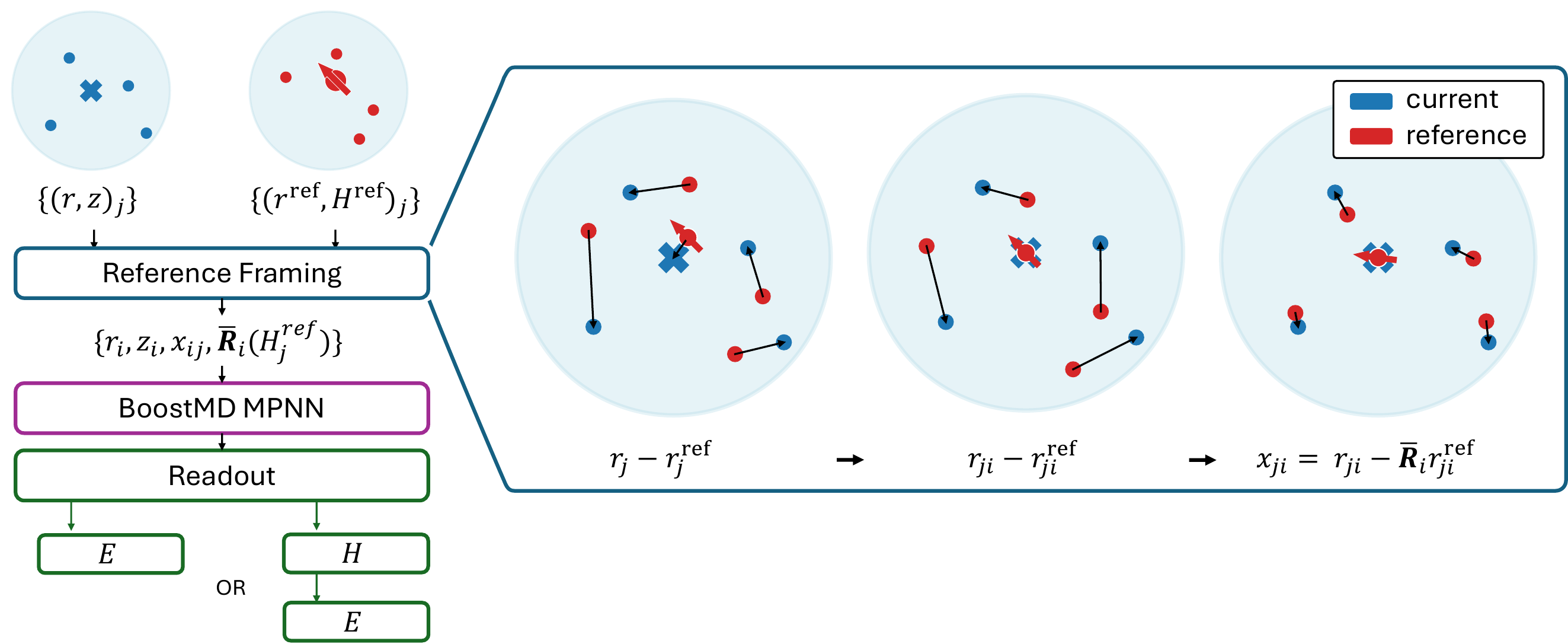}
    \end{subfigure}
    \hfill
    \vspace{0.05cm}
    \caption{\textbf{BoostMD architecture and reference framing} Showing the steps of BoostMD models, highlighting the reference framing step. The reference node features and reference positions are transformed to make BoostMD translationally and rotationally equivariant between steps as detailed in equation~\ref{eq:ref-frame-main}. The figures show the receptive field of an atom $i$ with neighbors $j$, showing both the current (blue) and reference (red) positions. The central red arrow represents an equivariant reference feature of the atom $i$, while the black arrows show the vectors associated with the label underneath each image. } 
 \vspace{-.25cm}
    
    \label{fig:reference-framing}
\end{figure}
  
  BoostMD models can be constructed using a variety of different architectures. We build on the equivariant MACE~\cite{batatia_mace_2023} blocks, which provide an efficient way to obtain many-body equivariant features. We use the tensor product notation ($\otimes$) introduced in Section~\ref{sec:background-notation} without explicitly writing the indices. Please see Appendix~\ref{app:equations-architecture} for a more detailed set of equations including all indices. 
 \vspace{-.2cm}

  \paragraph{Set Reference Frame} To be transitionally and rotational equivariant, the reference frame is transformed as visualized in Figure~\ref{fig:reference-framing}. Firstly, the origin is shifted to the central atom's current position $r_i$. Then, the reference is rotated by $\bar{\theta}$, as to minimize the L2 norm of displacements between the neighboring atoms' reference and current position, 
    \begin{equation}\label{eq:ref-frame-main}
        \mathbf{r}_{ji} = \mathbf{r}_j - \mathbf{r}_i, \qquad 
        \mathbf{r}_{ji}^\text{ref} = \mathbf{r}_j^\text{ref} - \mathbf{r}_i^\text{ref}, \qquad
         \mathbf{x}_{ji} = \mathbf{r}_{ji} - R_{\bar{\theta}_i}(\mathbf{r}_{ji}^\text{ref}), \\
    \end{equation}
    where $x_{ij}$ encodes the change in position and $r_{ij}$ encodes the current displacement. The optimal rotation $\bar{\theta}_i$ is efficiently computed using the Kabsch algorithm~\cite{kabsch_solution_1976} and ensures rotational equivariance. Further detail and its relevance for conservation of momentum can be found in Appendix~\ref{app:reference-framing}.
 \vspace{-.2cm}

    \paragraph{Equivariant message passing} We now construct a complete basis for equivariant functions of the change in positions $x_{ij}$, current positions $r_{ij}$ and the sender's reference node features $H_i^\text{ref}$, 
    \begin{equation}
    \hf_{ji} =  Y_{l_1}^{m_1}\left(\hat{\boldsymbol{r}}_{ji}\right)  \overset{|\mathbf{x}| |\mathbf{r}|}{\otimes} Y_{l_2}^{m_2}\left(\hat{\boldsymbol{x}}_{ji}\right) \otimes R_{\bar{\theta}_i}(\mathbf{H}_j^\text{ref}),
    \end{equation}
    where the first tensor product is learnable and depends on the magnitude of the displacement $|\mathbf{x}|$ and position vector $|\mathbf{r}|$ in a non-linear way. The second tensor product is also learnable, but only linearly mixes features in a way that maintains equivariance. 
    The edge features $\hf_{ji}$, can now be summed over all nearest neighbors $j$ within a cutoff to construct the BoostMD A-Basis,
    \begin{equation}
        \textbf{A}_i =  \left(\textstyle \sum_{j\in N(i)} \hf_{ij}^{(t)} \right),
    \end{equation}
    where the sum ensures permutational invariance of the energy. 
    
    The A-basis depends only on changes to atoms individually. To be able to capture effects that depend on two or more neighbors, we produce higher-body order features. To do this efficiently, we use the MACE~\cite{batatia_mace_2023} product basis to construct many-body messages, $\mathbf{m}$, from the two-body A-basis. This corresponds to taking tensor products of the A-basis with itself an arbitrary number of times, $\mathbf{m} = \mathbf{A} \otimes \mathbf{A} \otimes \dots \otimes \mathbf{A}, $ where all tensor products are learnable.  The messages now depend on, for example, the angles between two atoms.

    \paragraph{Readout} Finally, we use a readout MLP to map internal BoostMD features to (1) the energies $E_i$ directly or (2) the node features of the true MLFF at position $r_i$. For BoostMD with readout mode (2), the original reference's readout is used to predict the final atom-centered energy. While this Section focuses on the novel parts of the BoostMD architecture, the equations describing the entire architecture in detail, including indices explicitly, can be found in Appendix~\ref{app:equations-architecture}. The BoostMD layers are designed so that they can be used as layers in a message-passing neural network. While it is possible to run multilayer BoostMD models, for speed purposes, this is rarely of interest, and all presented results use single-layer models. 

 \vspace{-.1cm}

  \subsection{BoostMD at inference}

    In BoostMD, the full reference MLFF model is evaluated every $N$ steps, while a smaller, more efficient model operates in between. This new architecture can be used to accelerate molecular dynamics in various ways, each with different impacts on performance:
 \vspace{-.15cm}

  \begin{enumerate}
      \item \textbf{Serial evaluation of reference and BoostMD model} On a single GPU, the MD simulation pauses while the reference model is evaluated. After the reference features are calculated, the BoostMD model takes N steps and then waits for the next reference calculation. This approach trades some of the speedup for the benefit of directly incorporating the reference model's true forces into the MD run. For a model with speedup, s, the effective speedup is
      $$\frac{1 + (N-1)/s}{n}.$$
      \item \textbf{Parallel evaluation of reference and BoostMD model} Alternatively, the reference calculations could be performed on another GPU in parallel. This setup follows a producer/consumer pattern, where reference node features are continuously updated with the ground-truth MLFF and the BoostMD model uses the latest calculated reference features. The BoostMD model does not have to wait for the reference model evaluation, resulting in higher acceleration that is directly equal to s.
  \end{enumerate}

 \vspace{-.15cm}
    
    There are many more ways BoostMD models can be used at inference such as having adaptive step sizes based on observed errors or changes in receptive field. This broad design space underscores the potential of this approach for various applications.

 \vspace{-.2cm}

  \paragraph{Symmetries and Conservation Laws}
  The resulting trajectory from a BoostMD run does not conserve energy exactly. This is because the energy at a given position is dependent on the reference configuration. Consequently, it is possible to have a different energy predicted for the same configuration depending on the reference. 
  However, in between boost steps, which make up the large majority of steps, the method is energy, momentum and angular momentum conserving due to the reference framing. 
  Many popular machine learning force field architectures do not conserve energy~\cite{qu2024importance, liao2023equiformerv2}. The argument is that these architectures can learn energy conservation from data and that for finite temperature molecular dynamics the thermostat introduces random non-conservative force kicks even for energy-conserving force fields. Additionally, various methods in conventional atomistic simulation do not conserve energy, such as QM/MM calculations with adaptive QM regions~\cite{varnai_tests_2013}. It has been shown theoretically, that thermostats may ensure ergodic sampling at the correct temperature even in the presence of energy conservation violation~\cite{jones_adaptive_2011}. Empirically, we find that the lack of energy conservation is minimal relative to the noise added by the thermostat. This can most likely be attributed to the high energy-accuracy of BoostMD and the that the method is conservative in-between boost steps. Nevertheless, one should check for energy drifts and monitor the extent of non-energy conservation as the reference configuration changes.

 \section{Experiments}
 \vspace{-.2cm}

 In this section, we train BoostMD models to accelerate a foundation model for organic molecules, namely MACE-OFF23~\cite{kovacs_mace-off23_2023}. The force field has been shown to replicate the folding dynamics of peptides and accurately predict experimental observables relevant for drug discovery. We show how choices in the BoostMD architecture affect accuracy and performance, illustrating up to 8x speedup and stable dynamics.   

 \subsection{Dipeptide dataset: the building blocks of proteins}
 \vspace{-.15cm}

    \begin{table}[]
        \centering
        \begin{minipage}{0.75\textwidth}
            \centering
            \scalebox{0.9}{\begin{tabular}{ccccccc}
                \toprule
                \textbf{method} & \textbf{model size} & \textbf{$L^\text{ref}$} & \textbf{mode} & \textbf{speedup} & \textbf{E (meV)} & \textbf{F (meV/\AA{})} \\
                \midrule
                               & XS & 0 & feats & 8.4 & 1.30 & 64.5 \\
                \small BoostMD & XS & 0 & energy & \textbf{8.6} & 0.84 & 75.3 \\
                \small Model   & XS & 1 & feats & 2.3 & 0.64 & 57.6 \\
                               & XS & 1 & energy & 2.3 & \textbf{0.63} & \textbf{56.9} \\
                \midrule
                \small MACE    & XS\textsuperscript{+} & - & - & 5.6 & 1.80 & 121.1 \\
                \bottomrule
            \end{tabular}}
        \end{minipage}
        \hfill
            \begin{minipage}{0.24\textwidth}
            \centering
            \includegraphics[width=\linewidth]{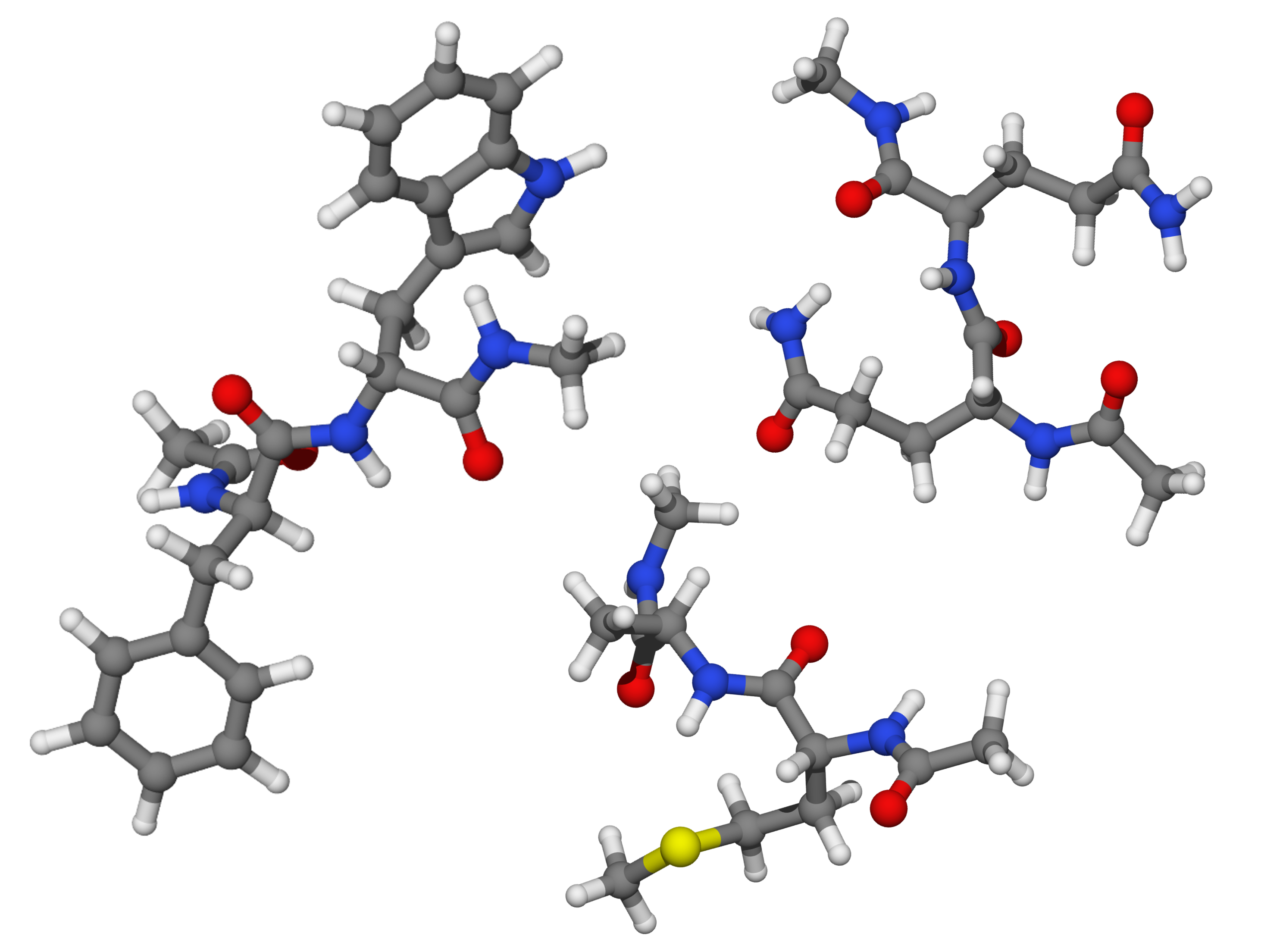} 
        \end{minipage}%
        \vspace{.3cm}
        \caption{\textbf{Comparing design choices on dipeptide dataset.} BoostMD models trained on a dipeptide MD dataset subselected from SPICE~\cite{eastman_spice_2023}. The energy per atom and force RMSE are computed with respect to the MACE-OFF23-medium ground truth. The reference model, MACE-OFF23-M, achieves an error of 0.85 meV/atom compared to DFT, higher than many of the boost models. Speedups are compared to the ground truth model. For details on hyperparameters corresponding to size (XS, XS\textsuperscript{+}) and the timing see  Appendix~\ref{app:model-hyper-params} and \ref{app:timing-experiments} respectively. }
        \vspace{-.5cm}
        
        \label{tab:res-dipeptides}
    \end{table}

 To assess the performance gains of various BoostMD models, we trained them on a dataset of dipeptides, generated by running molecular dynamics (MD) simulations using the MACE-OFF23 medium foundation model~\cite{kovacs_mace-off23_2023}. The training set dynamics are initialized from the dipeptide subset of the SPICE dataset~\cite{eastman_spice_2023}. We observe that single-layer BoostMD models provide speedups exceeding $8{\times}$, as shown in Table~\ref{tab:res-dipeptides} with low errors. With a single layer, the BoostMD model has a receptive field of 5\AA{}, in principle allowing for more efficient parallelization across GPUs using domain decomposition.

MACE-OFF23-M achieves an error of 0.85 meV/atom compared to DFT, with BoostMD models showing similar or lower accuracy relative to MACE-OFF23-M. We explored models, using only invariant reference node features ($L^{\mathrm{\text{ref}}} = 0$) as well as those incorporating equivariant features ($L^{\mathrm{\text{ref}}} = 1$). The inclusion of equivariant features improves accuracy at some computational cost. Additionally, we examined the two readout strategies from Section~\ref{sec:architecture}, finding that directly predicting energy changes is both marginally more accurate and faster than predicting changes in node features.

A natural alternative to training a BoostMD model would be to simply train a faster model from scratch.  Motivated by this, we trained a single-layer MACE model (XS$^+$), chosen to provide the same order of speed-up over the reference MACE-OFF23-M model as achieved by BoostMD.  However, we find that the energy and forces are less accurate than those achieved by BoostMD, highlighting the  advantages of biasing on previous features.


 \vspace{-.1cm}

  \subsection{Sampling of unseen Dipeptide}
 \vspace{-.15cm}

  Beyond low RMSE values, the true test of BoostMD is its ability to accurately sample equilibrium configurations at a given temperature. We found that BoostMD is stable during MD simulations, remaining robust even after \SI{10}{\nano\second} (\num{e7} \text{ steps}). Using molecular dynamics to sample configurations directly is restrictive in terms of the timescales that can be accessed. It is possible to accelerate sampling using meta-dynamics. Here, biases are introduced to sample less likely configurations in a controllable way, such that the probabilities of the unbiased simulation can be computed. In this Section, we test whether BoostMD models, trained on only MD samples, can accurately run meta-dynamics. This is a significant extrapolation as the model has not seen such unlikely (high energy) configurations during training. 
  
\begin{figure}[tb]
    \centering
    \centering
    \includegraphics[width=\textwidth]{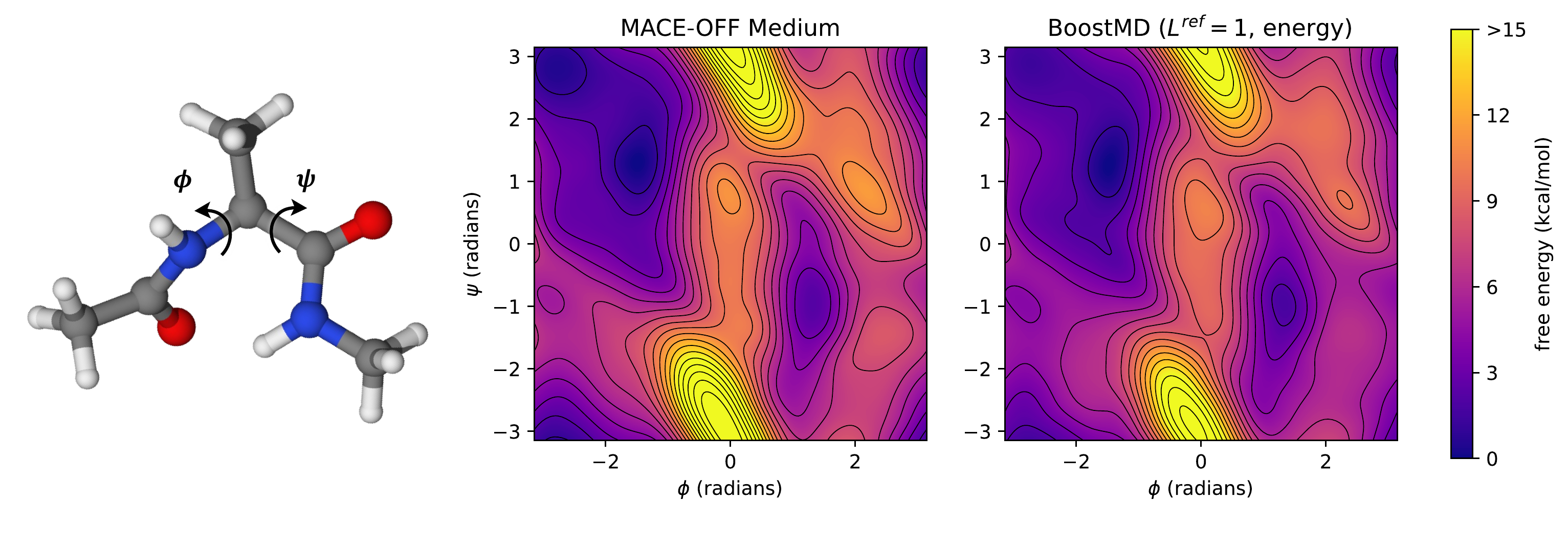}   
    \vspace{-0.7cm}
    \caption{\textbf{Free energy surface of unseen dipeptide.} Comparison of the samples obtained by running met-dynamics using the ground truth MACE-OFF model and BoostMD. The free energy of the Ramachandran plot, is directly related to the marginalized Boltzmann distribution $~\exp{[-F(\phi, \psi)/k_BT]}$, where $\phi, \psi$ are the dihedral angles marked in rendered molecule (left) . During BoostMD, the reference model is evaluated every 10 steps. Both simulations are run for 5~ns ($5 \times 10^6$ steps).}
    \label{fig:alanine-dipeptide}
    \vspace{-.3cm}
\end{figure}

  We find that when tested on an unseen molecule, alanine dipeptide~\cite{gfeller2007complex}, BoostMD accurately samples molecular configurations. Figure~\ref{fig:alanine-dipeptide}, shows the free energy, $F$, as a function of the backbone angles. The typical Ramachandran plot is directly related to the marginalized Boltzmann distribution $~\exp{[-F(\phi, \psi)/k_\mathrm{B}T]}$, where $T$ is the temperature. The strong agreement between the ground truth method (MACE-OFF medium) and BoostMD demonstrates that BoostMD can efficiently and accurately sample configurations, making it a valuable tool for long-timescale molecular simulations.

 \vspace{-.2cm}
 \section{Discussion}
 \vspace{-.2cm}

    In summary, we have developed BoostMD, a surrogate architecture that significantly accelerates molecular sampling with machine learning force fields (MLFFs). By leveraging highly descriptive node features computed at previous time steps, BoostMD predicts energies and forces at the current simulation step using smaller, faster models. This approach reduces computational costs, achieving up to an eightfold speedup with minimal loss of accuracy.

    Importantly, BoostMD maintains its performance on unseen systems, accurately sampling the Boltzmann distribution of dipeptides not encountered during training. This robust generalization makes it a valuable tool for efficient, long-timescale molecular simulations.

    Looking ahead, we plan to enhance BoostMD's capabilities by implementing concurrent training with the reference model, adopting adaptive inference schemes, and scaling training to larger datasets. We anticipate that BoostMD will enable routine, accurate, large-scale simulations, thereby accelerating the computational screening of biological and material compounds for desired properties, from catalytic activity to drug discovery.


\printbibliography[title={References}]



\newpage

{\huge\textbf{Appendices}}

\appendix

 \vspace{-.2cm}

\section{Detailed equations and BoostMD architecture}
\label{app:equations-architecture}

In this section we provide the details of the Boost MD architecture. The BoostMD (boost molecular dynamics) model predicts the node features, $\mathbf{h}_i$, and energy, $E_i$, of atom $i$ at position $\mathbf{r}_i$, given a set of reference node feature $\mathbf{H}^\text{ref}_j$ for a configuration with positions $\mathbf{r}^\text{ref}_j$, where $j$ is within the neighbourhood \( \mathcal{N}(i) \) of atom \( i \). This task can be approached with many different architectures. The first BoostMD model we present in this paper makes use of blocks from the \texttt{e3nn}~\cite{geiger_e3nn_2022} and \texttt{MACE}\cite{batatia_mace_2023} library. 

We start by motivating the choice of coordinates, ie reference framing. We then provide equations for the entire architecture.  

\subsection{Reference framing}
\label{app:reference-framing}

To ensure translational and rotational equivariance, the reference frame is transformed, as visualised in Figure~\ref{fig:reference-framing}. We refer to the sending and receiving nodes/atoms as $j$ and $i$ are the respectively.

Firstly,  we determine the displacement between atoms $j$ and $i$ for both the current position, $\mathbf{r}_{ij}$ and the reference position, $\mathbf{r}_{ji}^\text{ref}$, 
    \begin{equation} \label{eq:referencing-app}
        \mathbf{r}_{ji} = \mathbf{r}_j - \mathbf{r}_i, \qquad 
        \mathbf{r}_{ji}^\text{ref} = \mathbf{r}_j^\text{ref} - \mathbf{r}_i^\text{ref} \qquad
    \end{equation}

    This ensures translational equivalence, which guarantees conservation of momentum in between BoostMD steps and that the forces sum to zero. 

    To handle environment changes that correspond purely to rotations, parameterizing changes in displacement as \( \mathbf{r}_{ji} - \mathbf{r}_{ji}^{\text{ref}} \) is insufficient because rotational changes should directly map to rotated node features. To ensure exact equivariance, we compute the optimal rotation matrix $\bar{\mathbf{R}}_i$ that best aligns the reference configuration with the current configuration:

\[
\bar{\mathbf{R}}_i = \arg\min_{\mathbf{R}_i} \sum_{j \in \mathcal{N}(i)} \left\| \mathbf{r}_{ji} - \mathbf{R}_i \mathbf{r}_{ji}^{\text{ref}} \right\|^2.
\]

    The optimal rotation $\bar{\mathbf{R}}_i$ is efficiently computed using the Kabsch algorithm~\cite{kabsch_solution_1976}. The rotation is also applied to the reference node features and guarantees rotational equivariance between BoostMD steps. The change in positions, $\mathbf{x}_{ij}$, is hence defined as
    \begin{equation}
    \mathbf{x}_{ji} = \mathbf{r}_{ji} - \bar{\mathbf{R}}_i \mathbf{r}_{ji}^{\text{ref}}, \\
    \end{equation}
    
    \subsection{First layer-features} 
    
    Firstly we take a tensor product between the changes in displacement with learnable weights, defining the $\mathbf{X}$ basis:
    
    \begin{equation}\label{eq:app-xbasis}
    {X}_{ji}^{l_3 m_3} =  \sum_{l_1 m_1, l_2 m_2} C_{l_1 m_1, l_2 m_2}^{l_3 m_3} F_{l_1 l_2 l_3}\left(r_{ji}, x_{ji}\right) Y_{l_1}^{m_1}\left(\hat{\boldsymbol{r}}_{ji}\right) Y_{l_2}^{m_2}\left(\hat{\boldsymbol{x}}_{ji}\right),
    \end{equation}

    where $F_{l_1 l_2 l_3}$ is a multi-layer perceptron (MLP) of the inter-atomic distance $\mathbf{r}_{ji}$ and the size of the change in displacement $\mathbf{x}_{ji}$. Before being passed to the MLP, the distances are embedded using Gaussian basis functions. We now perform a tensor-reduced tensor product~\cite{darby_tensor-reduced_2023} with the reference features,
    
    \begin{equation}
    \mathbf{A}_{i, k l_3 m_3}^{(1)} = 
    \sum_{l_1 m_1, l_2 m_2} 
    C_{l_1 m_1, l_2 m_2}^{l_3 m_3} 
    \sum_{j \in \mathcal{N}(i)} 
    F'_{k l_1 l_2 l_3} \left( r_{ji}, x_{ji}, \mathbf{H}_{j, l=0}^{\text{ref}}  \right) 
    {X}_{ji}^{l_3 m_3} 
    \sum_{\tilde{k}, t} 
    W_{k \tilde{k} l_2}^{(t)} 
    D_{\bar{\mathbf{R}} i} 
    {H}_{j, \tilde{k} l_2 m_2}^{\text{ref}, (t)}
\end{equation}

    where $D_{\bar{\mathbf{R}} i}$ is the Wigner-D matrix corresponding to the previously determined optimal rotation, $\mathbf{H}_{j, \tilde{k} l_2 m_2}^{\text{ref}, (t)}$ are the reference node features of the $t$th layer of the reference model and $F'$ is defined similar to $F$. Summing over all neighbours $\mathcal{N}(i)$, we can now employ the density trick and take tensor powers of the $\mathbf{A}$-basis to create many-body features. Here we directly use the MACE~\cite{batatia_mace_2023} architecture (equ. 10) to construct the $\mathbf{B}$-basis,
    \begin{equation}
        {{B}}_{i, \eta_\nu k L M}^{(1)}=\sum_{\boldsymbol{l m}} \mathcal{C}_{\eta_\nu, \mathbf{l m}}^{L M} \prod_{\xi=1}^\nu \sum_{\tilde{k}} w_{k \tilde{k} l_{\xi}}^{(1)} {A}_{i, \tilde{k} l_{\xi} m_{\xi}}^{(1)}, \quad \boldsymbol{l} \boldsymbol{m}=\left(l_1 m_1, \ldots, l_\nu m_\nu\right),
    \end{equation}
    where $\nu$ is the correlation order and $\mathcal{C}_{\eta_\nu, l \boldsymbol{m}}^{L M}$ are the generalised Clebsch-Gordan coefficients. For more details please see the MACE~\cite{batatia_mace_2023} paper directly. The $\mathbf{B}$-basis is used to update the node features, 
    \begin{equation}
    {m}_{i, k L M}^{(1)}=\sum_\nu \sum_{\eta_\nu} W_{z_i k L, \eta_\nu}^{(1)} {{B}}_{i, \eta_\nu k L M}^{(1)} , \qquad   {h}_{i, k L M}^{(2)} =\sum_{\tilde{k}} W_{k L, \tilde{k}}^{(1)} {m}_{i, \tilde{k} L M}^{(t)},
    \end{equation}
    where $\mathbf{h}$ are the internal BoostMD features. The architecture for the next layers of the BoostMD model are identical to that of MACE~\cite{batatia_mace_2023}. While it is possible to run multi-layer BoostMD models, for speed purposes this is rarely of interest. Consequently, all presented results are performed with single layer models. 

    \subsection{Readout} 
    
    Finally, BoostMD can either be used to (1) directly predict changes in energies $E_i$ or (2) changes in the node features of the true MLFF $\mathbf{h}_i$ at position $\mathbf{r}_i$. Assuming the BoostMD model has $t'$ layers, the final energy depends on the readout mode:
    \begin{enumerate}\label{eq:readout-energy}
        \item \textbf{Direct energy readout} The reference model has an energy associated with each of its layers $t$. We perform a readout on each of the BoostMD layers $t'$ for each reference layer energy $t$,
        \begin{equation}
E_i^{(t)} = E_i^{\text{ref}, (t)} + \sum_{t'}\texttt{MLP}^{(t')}( \mathbf{h}^{(t')}_{i}) \sum_j|x_{ij}|,
\end{equation}

    where the multiplicative factor $|x_{ij}|$, ensures that when there is no change in environment, then there is no change in predicted energy.  

    \item \textbf{Reference node feature readout} Here we predict the changes to the reference's models node features. 
    \begin{equation}
        \mathbf{h}^{(t)}_{i} = R_{\bar{\theta}_i}(\mathbf{H}_{i,k}) +  \sum_{t'}\texttt{MLP}^{(t')}( \mathbf{h}^{(t')}_{i}) \sum_j|x_{ij}|
    \end{equation}

    The original readout of the reference model $\mathcal{R}_t$ can then be used to compute the energy as
    \begin{equation}\label{eq:readout-nfeats}
    E_i^{(t)}=\mathcal{R}_t\left(\boldsymbol{\mathbf{h}}_i^{(t)}\right).
    \end{equation}

    \end{enumerate}

The final total energy is simply a sum over layer energies and atoms, $\sum_{t,i} E_i^{(t)}$. Forces can be determined using backpropagation. As previously mentioned, for speed reasons all experiments are performed with a single layer BoostMD model.

\section{Node feature and their properties}
\label{app:node-feature-properties}

 \begin{figure}[tb]
    \centering
    \begin{subfigure}[b]{0.85\textwidth}
        \centering
        \includegraphics[width=\textwidth]{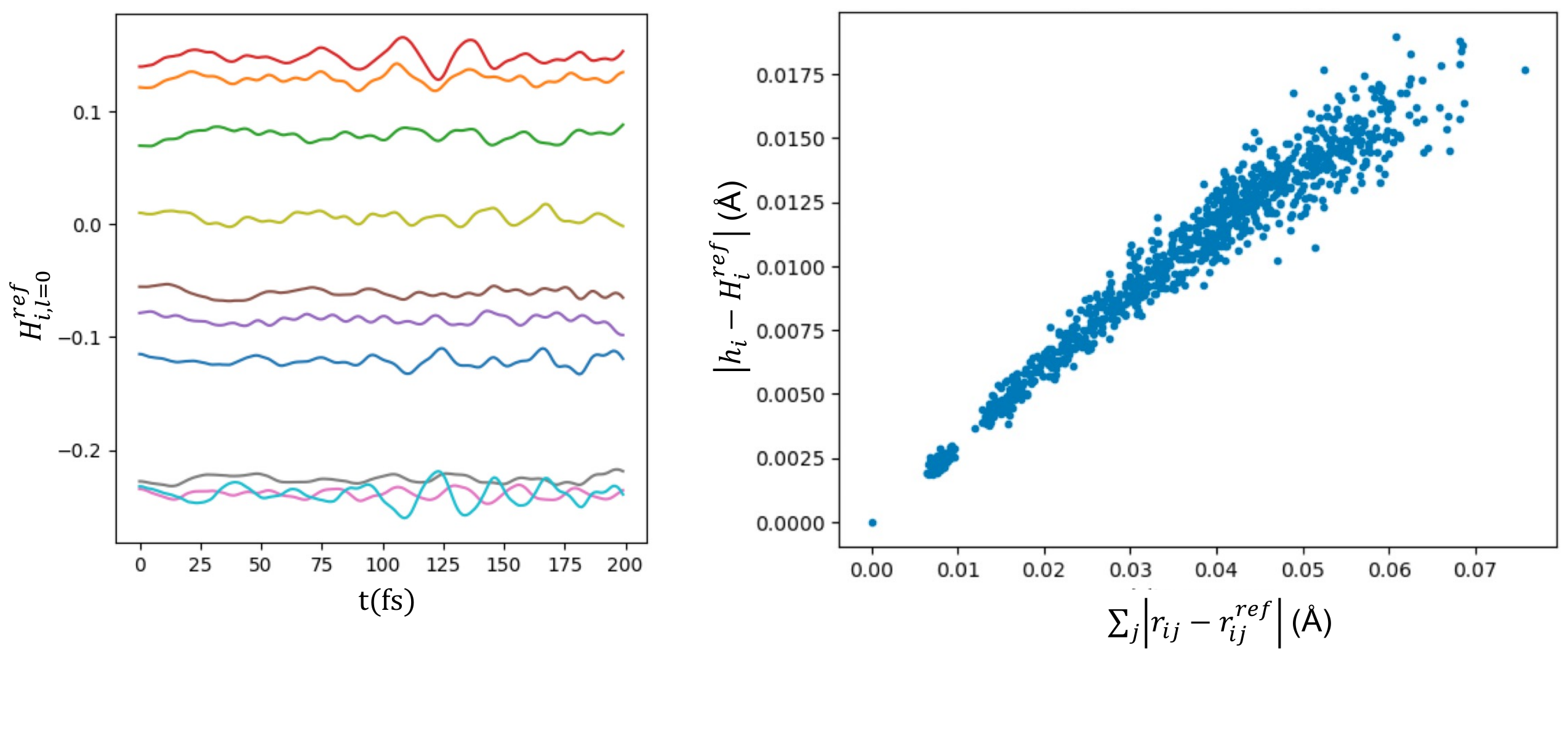}
    \end{subfigure}
    \hfill
    \vspace{-.5cm}
    \caption{\textbf{Node feature properties} Showing the fluctuations of the reference node features as a function of simulation time (\textbf{a}) and the change in node features as a function of change in position inside the atomic environment. } 
    \label{fig:node-feats}
\end{figure}


\section{Details on Experiments}
\label{app:experiments-details}

\subsection{Model hyper-parameters}
\label{app:model-hyper-params}

The aim is to produce a fast architecture while maintaining sufficient accuracy. We restrict ourselves to a single layer BoostMD model as this not only reduces the model cost, but would also allow for more parallelisation. For big simulations, local models can be partitioned into smaller overlapping regions and evaluated in parallel on multiple GPUs. The size of the overlap needed depends on the total receptive field. A single layer reduces the receptive field from 10\AA{} to 5\AA{} significantly increasing its prallelisation abilities. 

For the BoostMD models, we define the XS size as a single model, with correlation order 2, compared to 3 for the reference foundation model. Furthermore, the maximum order of the spherical harmonics is set to 2, compared to 3 for the MACE-OFF-M model. We use 125 channels for the BoostMD model. As the BoostMD architecture requires one additional edge tensor product in equation~\ref{eq:app-xbasis}, comapred to the MACE architecture, a MACE model is faster than a BoostMD model with the same hyperparameters. For a fair comparison, we hence give the compared MACE model additional flexibility by increasing the number of channels to 256 (XS\textsuperscript{+}). As visible in Table~\ref{tab:res-dipeptides}, even with this added flexibility the MACE model underperforms a BoostMD model, which depends on previously computed node features. All experimental results do not use the rotational reference framing during training or inference, due to the associated computational cost. 
\subsection{Timings}
\label{app:timing-experiments}
Timing measurements were performed on a system consisting of 3BPA molecules, with a total of 3,375 atoms. This large system size is representative of typical biomolecular simulations and minimizes the impact of PyTorch overhead on the measured runtime. The reported timings are averaged over 100 evaluations, following an initial warm-up period of 50 evaluations, and were performed on an NVIDIA A100 GPU.
\subsection{Free energy surfaces}
The free energy surfaces of Figure~\ref{fig:alanine-dipeptide} were calculated with the PLUMED package~\cite{bonomi2009plumed}, using both backbone angles as collective variables. The N-Acetyl-dl-alanine methylamide molecule was taken from Reference~\cite{gfeller2007complex} and has PUBChem CID 141892. This exact molecule is not part of the training dataset.

\end{document}